\documentclass[a4paper]{jpconf}
\pdfoutput=1
\usepackage{graphicx}
\usepackage{xcolor}
\usepackage{amsmath}
\usepackage{placeins}
\usepackage{hyperref}
\usepackage{subcaption}

\graphicspath{{Figs/}{./}}

\def\orcid#1{\kern .08em\href{https://orcid.org/#1}{\includegraphics[keepaspectratio,width=0.7em]{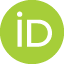}}}

\begin{document}
\title{How can we explain last UHERC anisotropies?}

\author{D.\,Fargion$^{1}$\orcid{0000-0003-3146-3932}, P.G.\,De Sanctis Lucentini$^{2}$\orcid{0000-0001-7503-2064}, M.Y.\,Khlopov$^{3,4,5}$\orcid{0000-0002-1653-6964}}

\address{$^{1}$Physics Department, Rome University 1, P.le A. Moro 2, 00185, Rome, Italy}

\address{$^{2}$ Gubkin University, Leninskiy prospekt 65, Moscow, 119991, Russia}

\address{$^{3}$Center for Cosmoparticle physics Cosmion, National Research Nuclear
University ”MEPHI”, Moscow 115409, Russia}
\address{$^{4}$Institute of Physics, Southern Federal University, Stachki 194 Rostov on
Don 344090, Russia}
\address{$^{5}$Virtual Institute of Astroparticle Physics, Paris 75018, France}

\ead{daniele.fargion@fondazione.uniroma1.it}

\begin{abstract}
We consider the recent results on UHECR (Ultra High Energy Cosmic Ray), clustering,  composition, distribution in the sky, 
    from the energy of several EeV with the dipole anisotropy up to the highest ones.  We have suggested  since 2008  and we reconfirm here that UHECR at $40-70$ EeV are mostly made by light and lightest nuclei. The remarkable Virgo absence and the  few localized nearby extragalactic sources as CenA, NG 253, and M82 may be well understood by the lightest nuclei fragility and opacity within Mpc distances.   We comment also on the role of a few  galactic UHECR sources at ten EeV that may be  partially feeding  the Auger dipole UHECR anisotropy. 
        The recent anisotropy in the UHECR spectral composition of lightest and heavy nuclei, outside and along the galactic plane, could also be a first confirmation of our previous claims (2012).
         The  interplay of the heavier and most energetic  UHECR  galactic nuclei  with the mainly local (Mpcs) extragalactic signals ruled by lightest nuclei, seems to fit the main pieces of UHECR puzzle.        
\end{abstract}

\section{Introduction: The Virgo Absence and the two Hot Spots}

The Ultra High Energy Cosmic Rays (UHECR) after two decades of data are, probably, at a first stage of understanding. Because of their nature, the hadron charged Cosmic Rays (CR) are bent and smeared by magnetic fields both  from near, galactic, and by far cosmic  distances. The lepton CR, electron pairs, instead, lose very fast their energy and have a very local (parsecs) origin, bounded within lowest energy (TeVs-PeVs) ones. However, the highest energetic CR, nucleon or light nuclei, should be able to point back to their source with a good directionality thanks to their high rigidity. Their heavier nuclei  directionality  may unfortunately be anyway smeared by their largest charges.     Therefore the understanding of UHECR is a hard combined puzzle of  energy, composition, magnetic fields and volumetric source distribution and nature. Anisotropy maps, event rates, UHECR compositions and the wide astrophysical data maps have to be all tuned at once.

In last two decades, largest area detectors for UHECR after Fly's Eye, as the AGASA \cite{AGASA(1998)}, the  High Resolution Fly's Eye (HiRes), the Telescope Array (TA), the Pierre Auger Observatory (Auger), they were able to record several hundreds above $6\cdot10^{19}$ eV UHECR events, near or well above the GZK cutoff \cite{GZK(1966),GZK(1966)b}.

Moreover, the nucleon or the light nuclei UHECR disruption and fragmentation (due to the photo pion,the GZK cut off, or to the photo nuclear dissociation, both caused by the cosmic radiation interaction) make the UHECR proton confined into a small (hundred Mpc, the GZK cut off) or even, lightest nuclei, within a  very local (few Mpcs) Universe. The heaviest CR like iron, Ni, Co  are  much bent by their large charges and cannot be easly disentangled in the sky.  Heaviest ones,  may be even turn in spirals,  possibly captured inside the galaxy.  Proton, the most popular courier for last decades,  may fly nearly straight (clustering within a few  degrees,$\simeq 3^{\circ}$ at $ 6 \cdot 10^{19}$ eV). But since a decade there are not in AUGER, in terrestrial South, or in TA (Telescope Array), in the North, such characteristic narrow clusterings of observed UHECR. 
Lightest UHECR nuclei (He,D,Li,Be..) directions are  little smeared,  still enough collimated, up to $\simeq  10^{\circ}$ at $6 \cdot 10^{19}$. That is a typical size  as the first observed UHECR anisotropy (the twin Hot Spot) in AUGER and later in TA. Moreover lightest UHECR nuclei (He like) are constrained much within a few Mpc Universe by photonuclear distruption and fragmentation. This opacity is avoding any ten-tens Mpcs distance: this is an  the ideal filter or an opacity toward the Virgo cluster, explaing its amazing and puzzling absence both in  AUGER and TA sky.  Additional role of few heaviest nuclei at highest UHECR energy had been also considered \cite{fargion2012apart}.  Most of the frame and details of the present article,  are included in an earlier contribute \cite{fargion2018POS} and references therein.

Let us remind first that the charged nuclei or nucleon in CR are easily deflected by the astrophysical (galactic and intergalactic) magnetic fields via Lorentz forces. Therefore, heavy charged nuclei ($A>> 4$) CR are soon bent and turn,  loosing their primordial  directionality. However, the CR of the highest energies above $10^{19}$ eV (tens EeV), if protons, suffer less and less from deflection promising a geometrical connection with their original sources. Eventual Ultra High Energy Cosmic Ray (UHECR) neutrons (born by photopion interactions) being neutral are un-deflected but they are bounded by their instability within a very narrow cosmic radius: $\simeq Mpc \frac{E_{n}}{10^{20} eV}$. Many bounds exist on EeVs neutrons, but they might also play a very minor galactic role.

 First  we noted the remarkable absence of Virgo in the UHECR clustering since 2007 \cite{PAO(2007)} discover, a clustering expected for UHECR proton composition. A first low clustering rised also around Cen A nearest AGN.
   The heavy iron nuclei (considered by most as ideal courier) could not cluster as a narrow angle  as this first Auger Hot spot size. Therefore we considered the lightest nuclei as the solving candidate, being well filtered and opaque from far Virgo distances, but transparent from Cen A, bent by a needed size angle. Five years later it became quite agreed that UHECR at such energies are mainly light nuclei (or lightest ones),  \cite{PAO(2017b)} (see also therein, the note at page 26, and their references $83-87$), by their average slanth depth of the shower maximum.
 We noted the characteristic bending (or deflection) angle expected by random or coherent deflection by Lorentz forces \cite{Fargion(2009b)}, mainly due to galactic magnetic fields. These angles agree well with hot spot smeared angle size.
 We remind  our consequent foreseen lightest nuclei fragmentation from Cen~A UHECR above  $ 40-60$  EeV, into fragment around twenty EeV ones \cite{Fargion(2009b)}. Two years  later Auger observed such fragments in multiplet tails near 20 EeV energy. These twin tails around Cen A had been observed as well as a similar one along LMC and SMC  multiplet published in a later article \cite{PAO(2011)}. This multiplet near LMC and SMC suggest a minor galactic role of UHECR. The same conclusion arised from the UHECR dipole anisotropy discussed later.

\begin{figure}[t]
\begin{center}
\includegraphics[width=0.78\columnwidth]{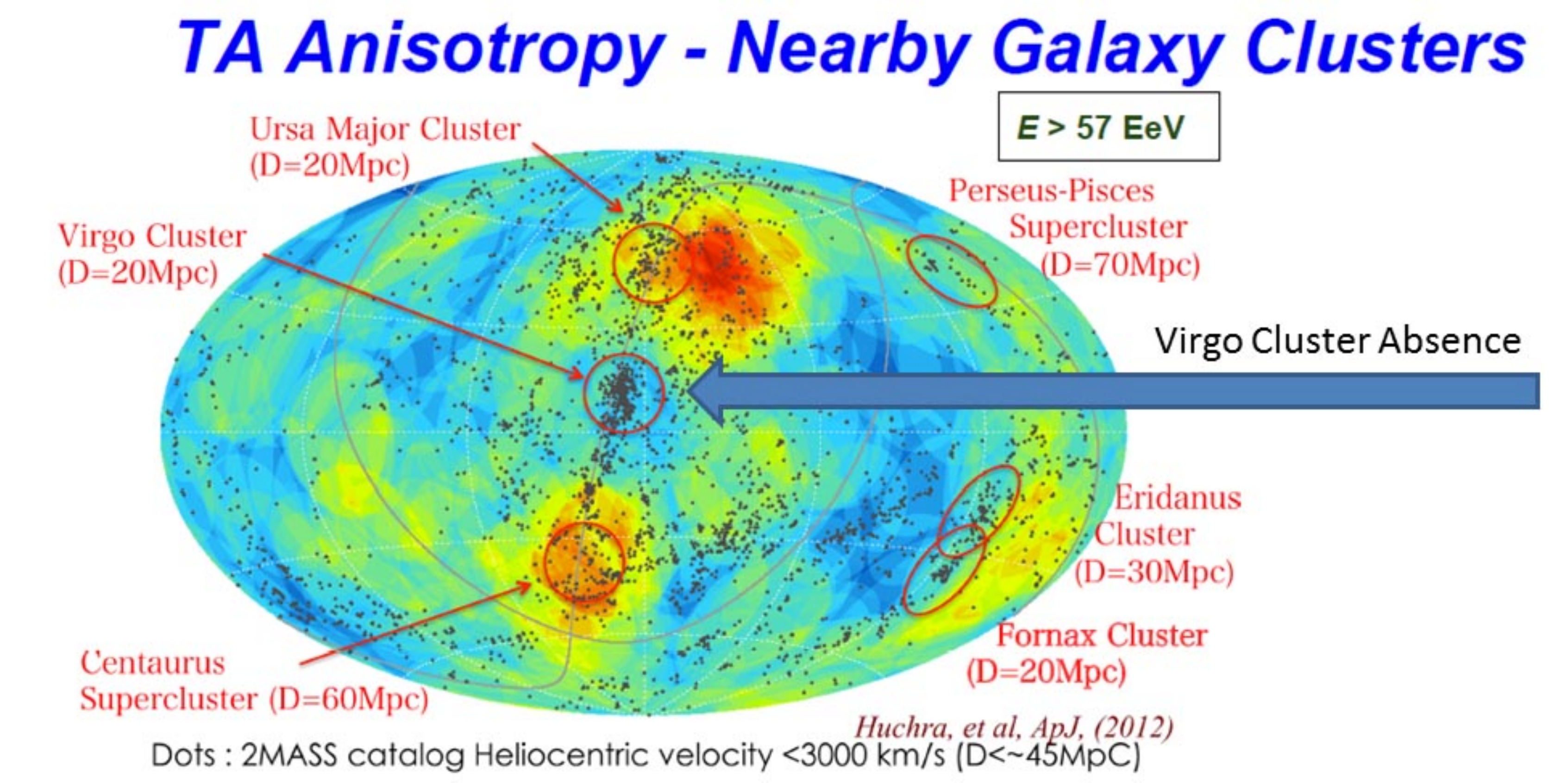}
\caption{The combined  map, in equatorial coordinates, based on the 2015 Auger and TA presentations where the two red hot spot UHECR clustering signals, North for TA and South for Auger,  are showing the main UHECR dense event rate. The dark dots are the main galaxies in nearby GZK volumes\cite{Huchra2012RedShift} expected by any proton UHECR courier. This Virgo signature is missing, overlapping the abundant Virgo galaxy cluster sources: their heaviest star mass concentration is within the GZK volume. To explain this absence   we imagined and suggested a kind  of a filter. We claimed that UHECR were mostly lightest nuclei, too fragile to survive and to reach us above a few Mpc because the cosmic big bang thermal bath. Indeed Virgo lay at 20 Mpc while the lightest nuclei interaction distance it is below a few Mpc. Cen A as well as M82, NGC 253, are only few Mpc. Only a decade later the light nuclei UHECR have been confirmed  (see note at page 26 of \cite{PAO(2017b)}) by average slanth depths of the air-shower maximum.}
\label{Fargionfig1}
\end{center}
\end{figure}

Incidentally and historically, let us also to remind the earliest Fly's Eye 300 EeV event from 1995. Its  huge energy, over GZK cut off  \cite{GZK(1966),GZK(1966)b}, its arrival direction out of any known important nearby source, led us to consider an UHE neutrino courier scattering onto relic light (about $eV$ mass) ones. These ZeV UHE neutrino might overcome the GZK cut off, a very severe one for any nucleon or nuclei. These UHE neutrino scattering (by resonant Z boson production) onto the relic cosmic warm  ones (at eV mass)  could allow its secondariy hadron fragments to reach us from unexpected distances, overcoming any GZK cut off and explaining  earliest and also recent puzzling correlations with very far AGN sources. Indeed  recent mild UHECR clustering along 3C 454 and a very last and energetic TA (Telescope Array) event from unknown sources, if confirmed, might need such an extreme solution. Eventually, within a less exotic solution, the puzzle requires a highest UHECR energy, heavy nuclei, bent and turn from possibly nearby, even galactic \cite{fargion2012apart}, sources.

 Finally the large unexpected anisotropy ($6.5\%$) of ($8-10$) EeV UHECR energy have shown since 2017 a remarkable ($>5$ sigma) signal showing an amazing UHECR dipole anisotropy, enhanced in last years: pointing neither to the galactic center nor to the nearby Super Galactic Plane. We suggest here a different view based on a mixed galactic and a very local extragalactic component made by  very few candidate sources.

\subsection{UHE ZeV neutrino scattering to solve UHECR and TeV puzzles?}
A cornerstone for a neutrino mass new physics might be linked with the 3C~454 AGN clustering of UHECR events (see Fig.\ref{Fargionfig8} in this paper and also Fig. 5, arrow A in \cite{Fargion2017NarrowClustering}).
Also Mrk~421 and PKS~0208-512  may play such a role in additional two hot spots.
Such a far source (3C~454) is near $2.5$ Giga-parsec distance, almost 50 times above the most penetrating proton UHECR: there is no way to reach directly to us overcoming the exponential ($\simeq e^{-50}$) GZK cutoff.
Therefore, the possible role of a relic mass neutrino of nearly one eV mass  may play the role of a calorimeter able to capture a different courier (a ZeV anti-neutrino) in order to explain this puzzling connection \cite{Fargion.et.al.(1999)}, \cite{Fargion2007Splitting}.
More in detail the AGN may shine UHECR also at ZeV energies for neutrinos that cross the Universe with negligible absorption  and hit in a wide relic neutrino mass halo (several or tens Mpc wide) making a Z boson:
such forming UHE Z resonance, its decay in hadrons (involving nucleon-antinucleon pairs) is the final UHECR message reaching us.
The consequent presence of UHE EeV and ZeV tau neutrinos may also lead to upward 
tau air-showers~\cite{Fargion(2000), Fargion.et.al.(2004), PAO(2009), PAO(2008)}.
The very recent and popular sterile neutrino at $1.6$ eV mass might be incidentally well tuned to produce UHECR at a peak of $5 \cdot 10^{19}$ eV \cite{Fargion2007Splitting}.


We proposed \cite{Fargion.2018} and confirm here that this clustering, if growing and being confirmed, might be associated and originated by a far beamed AGN, whose ZeV neutrino are scattering onto a relic cosmic neutrino background (with $\simeq 1-2$eV mass) via Z boson resonance (the so called Z burst model) \cite{Fargion.et.al.(1999), Fargion2007Splitting, Fargion2004Colaiuda, Fargion2017NarrowClustering}. Indeed the UHE ZeV neutrinos, emitted by far AGN do not interact with the cosmic photons and do not suffer of any GZK cutoff as photons. A ZeV UHE neutrino does scatter  with relic neutrinos halos (galactic or local) by a tuned Z boson resonance, whose peaked cross section \cite{Fargion.et.al.(1999)} makes their detection possible.
The UHE  Z boson decay in pions, nucleons and anti nucleons might be the observed as UHECR.   
The additional EeVs hard gamma secondaries, their brightening overcoming any  GZK cut off, is also leading to a secondary TeV gamma showering. Such TeVs gamma may overcome also the severe cosmic infrared background opacity.
 \cite{Fargion2004Colaiuda,Fargion2017NarrowClustering}. 
 
 This ability to Z resonant neutrinos to feed TeVs signals might solve  the very recent and amazing hard gamma  event, by the GRB221009A,  whose $18 $ TeV event, that had been observed by LHAASO on 9th of October 2022. The GRB221009A is at cosmic distance $z= 0.15$ and a tens TeV photon cannot fly across the IR radiation without more than a million times suppression. Therefore the same UHECR by ZeV courier may also feed such Tens TeV amazing signal.
 
 These relic neutrinos target in  hot dark halo, might be by a mass  about $0.4-0.1 eV$. Also a neutrino mass by an exciting (but still controversial) sterile neutrino (claimed  tuned at $1.6$ eV) may solve the model. Similar arguments apply to  the AGN Mrk~421, (at redshift:$ z=0,0308$), shining around the second UHECR hot spot at the north sky, observed  by TA.
 Finally  Lightest UHECR fragments, ZeV neutrino,  may provide, as GZK ones, source for lower energy neutrino signals (at ten-hundreds PeVs) at smaller fluency, better observable in spectacular upward tau air-shower signals \cite{Fargion(2000), Fargion.et.al.(2004), PAO(2009), PAO(2008)} observable in future mountains, balloons or satellite detectors.

.


\section{Lightest UHECR deflection  in random or coherent path walk and their delay}
The magnetic fields are able to deform the charged CR trajectory in an almost predictable way. For UHECR with energies above few tens EeV, the deflection occurs from far cosmic distances and by galactic near fields. The galactic magnetic field is mainly correlated with the galactic arms. The flight across
our galactic plane must suffer of a random bending (or deflection) (if the trajectory it is skimming on our galactic planes, as for the case of Cen A sources, up and down respect the galactic plane) \cite{fargion2011coherent}. The galactic field plays a minor role if the path of UHECR is far from the galactic plane.
 Then the deflection may be simple a coherent one, not a random one \cite{fargion2011coherent}. This may be the case of the AGN  M82 UHECR trajectories that are coming from the pole of our galaxy.
 While the random walk orthogonal to the galactic plane may lead to an \textit{up-down} bending, orthogonal to the galactic plane disk, as for Cen A, the coherent bending might be the cause of an asymmetric disposal of the UHECR cluster of events with respect to the optical or gamma source candidate.
 In  case of UHECR lightest nuclei, the deflection angle is scaling with the atomic number Z, respectively
($Z_{\mathrm{He}}=2$, $Z_{\mathrm{Li}}=3$, $Z_{\mathrm{Be}}=4$, $Z_{\mathrm{B}}=6$)
because of the extragalactic and mainly because of the galactic deflection angle. Let us evaluate therefore both the two main contributions:
$$
\alpha^{gal}_{\mathrm{He}}\simeq\\
15.5^\circ\left(\frac{Z}{Z_{\mathrm{He}}}\right)\left(\frac{E}{6\cdot10^{19}\,\mathrm{eV}}\right)^{-1}\left(\frac{D}{20\,\mathrm{kpc}}\right)^{1/2}\left(\frac{d_c}{\mathrm{kpc}}\right)^{1/2}\left(\frac{B}{3\,\mu\mathrm{G}}\right)
$$

$$
\alpha^{ex}_{\mathrm{He}}\simeq\\
3.28^\circ\left(\frac{Z}{Z_{\mathrm{He}}}\right)\left(\frac{E}{6\cdot10^{19}\,\mathrm{eV}}\right)^{-1}\left(\frac{D}{4\,\mathrm{Mpc}}\right)^{1/2}\left(\frac{d_c}{\mathrm{Mpc}}\right)^{1/2}\left(\frac{B}{\mathrm{nG}}\right)
$$

which leads to a total of 
$\alpha^{ex}_{\mathrm{He}}+\alpha^{gal}_{\mathrm{He}}\simeq18.7^\circ$,
in good agreement with the observed Hot Spot spread angle size. The delay time is mostly due to the extra galactic magnetic field; thus the time delay between UHECR after its photon direct one, can be evaluated as:
$$
\Delta\tau_{\mathrm{He}}\simeq\\
2666\left(\frac{Z}{Z_{\mathrm{He}}}\right)^2\left(\frac{E}{6\cdot10^{19}\,\mathrm{eV}}\right)^{-2}\left(\frac{D}{4\,\mathrm{Mpc}}\right)^{2}\left(\frac{d_c}{\mathrm{Mpc}}\right)\left(\frac{B}{1\,\mathrm{nG}}\right)^2\,\mathrm{yr}
$$

Such a short cosmic delay allows a nominal  Cen~A (as well as the nearest M82, NGC~253) AGN or starburst galaxies to shine as
UHECR while being still active in optical, gamma, or X-ray, or radio band. On the contrary alternative proposals of much distant AGN \cite{Kampert(2016)}, \cite{Deligny(2017)} sources,  have a time of flight (from a hundred Mpc distances) that should suffer from a huge (million years) delay (respect to the direct gamma photons). Therefore, most far UHECR are very possibly uncorrelated \cite{Fargion.2018} with their present gamma observations \cite{Joshi(2018)}. The main message in conclusion is that the lightest nuclei as Helium might be the first and the main responsible courier of UHECR, source of the hot spot clustering observed by TA and Auger, in the last decade.
The UHECR source candidates and their coordinates will be described in the final sections. In following maps  we updated the two spot clusterings in celestial coordinates with their underlined (published in detail, 2015) UHECR event map.
The map will help us to better disentangle the UHECR sources \cite{2015arXiv151208794F}, adding also the last (updated but averaged, 2018) anisotropy maps \cite{PAO(2017)} in the last section.


\section{A dipole originated within a local and-or galactic space?}

  Here we address mainly to the very recent dipole anisotropy found by Auger at 10 EeV with an unexpectedly large ($6.5\%$)
weight \cite{PAO(2017)}. Its location in the sky and its statistical relevance suggest, within a cosmic noise,  a combined, but ruling galactic contribute (Vela, Crab, LMC, SMC)  as well as the nearby NGC 253 sourece,  a small nearby extra galactic presence. Note in the following dipole maps the shadowed area, where the UHECR exceed the background: this area it is not pointing to our galactic center, nor to nearest cosmic denser sky (as Virgo cluster or the great attractor one). The dipole is pointing versus the anti-galactic center where the bright Crab is located and where the galactic Vela, LMC, SMC sources (as well as the nearby Fornax~D, NGC~253) are also shining possibly as main candidate sources. Therefore, the latter sources may be mainly these galactic ones, offering an explanation  of the remarkable high UHECR over-abundance, as discussed in the final section.
Indeed, the other known dipole adimensional anisotropy, as the cosmic black body one it is a much smaller ($2  \cdot 10^{-3}$) one.
We claimed \cite{Fargion.2018} and confirm  that there are not any known extragalactic candidates able to fit the dipole anisotropy \cite{PAO(2017)}; the galactic origin candidacy, made by the above few sources, is probably the best one also with an AGN NGC~253 extragalactic component.
In particular, the random bending (or deflection) at 10 EeV for an Helium like nuclei is spread in a reasonable agreement with the observed dipole size $\pm 90^{o}$:
\begin{multline}\label{DIPOLE}
\alpha^{DIPOLE}_{\mathrm{He}}\simeq
93.0^\circ\left(\frac{Z}{Z_{\mathrm{He}}}\right)\left(\frac{E}{10^{19}\,\mathrm{eV}}\right)^{-1}\left(\frac{D}{20\,\mathrm{kpc}}\right)^{1/2}\left(\frac{d_c}{\mathrm{kpc}}\right)^{1/2}\left(\frac{B}{3\,\mu\mathrm{G}}\right)
\end{multline}

Moreover, the Auger mass composition measurements at 10 EeV favour a dominant He-like (lightest) nuclei, a partial light ones (possibly N, but also Be, B may fit)  and the near absence of proton or iron nuclei.

The additional highest energy (above 80 EeV) events might be ruled by heaviest Ni-Co-Fe nuclei, as discovered by recent \cite{Mayotte2022Aq} composition spectra map, asymmetry mostly reaching (as heavy nuclei) from the galactic plane, and reaching  from  outside the galactic plane (as lightest ones). 
     Therefore nearest sources,  even Crab, Vela and Geminga ones,  may also contribute by such heavy nuclei to the observed dipole anisotropy.  This local source pollutions and  asymmetry is somehow confirmed or tuned  with an additional recent observed asymmetry between North (TA) and South (AUGER) spectra at their maximal energetic edges  \cite{plotko2022indication}. Lightest nuclei are polluting from outside the galaxy  and may reach us from M 82, NGC 253, LMC,SMC and even M31, our nearest Andromeda companion galaxy within Mpc volumes.



\subsection{Cen A He UHECR photo-dissociation and their 20 EeV twin fragments}

We have suggested \cite{Fargion(2009b)} that  fragments had to arise and be detectable as multiplet events (from half to a fourth of the GZK energies) from Cen A.  Indeed they have possibly been observed by Auger in the 2011 at 20 EeV.

In the following map, the Auger results from 2018 regarding the anisotropy of CR with E $> 39$ EeV \cite{PAO(2018)} are the background of these interesting narrow multiplets of clustering, as explained in the caption of Fig.~\ref{Fargionfig6}.

Their  twin clustering  are suggesting that the deflection is ruled from Cen A  by the random galactic spiral alternate magnetic field lines. The one and a half deflection of the $20$ EeV multiplet of the fragment events (respect harder ones) is larger than the $6 \cdot 10^{19}$ eV highest energy clustered events around Cen A, because their energy is nearly a third of the highest ones (He-like), but their charge is nearly half (Deuterium like)


\subsection{Source candidate and locations for UHECR clusterings}

\begin{figure}[h]
  \mbox{}\hfill  
  \begin{subfigure}[t]{.48\linewidth}
    \centering\includegraphics[height=5cm, width=\linewidth]{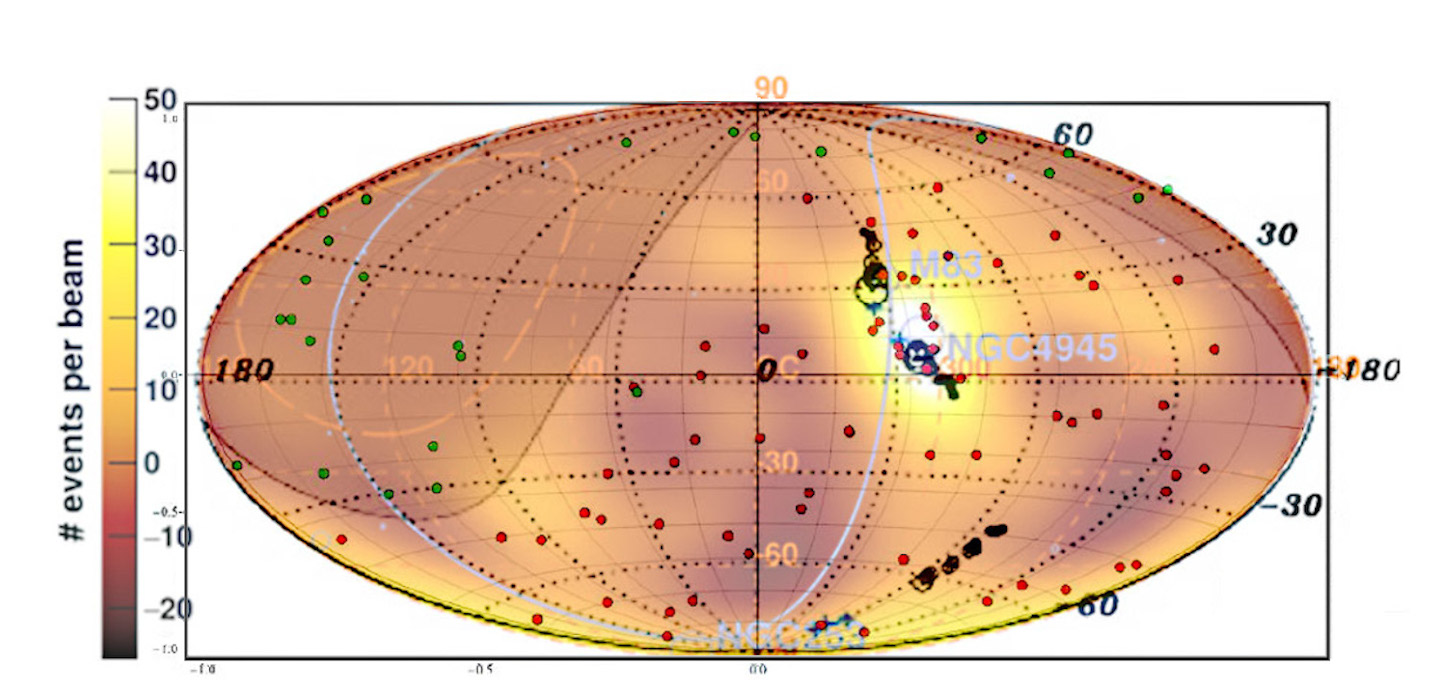}
    \caption{\label{Fargionfig6}}
  \end{subfigure}\hfill
  \begin{subfigure}[t]{.48\linewidth}
    \centering\includegraphics[height=5cm, width=\linewidth]{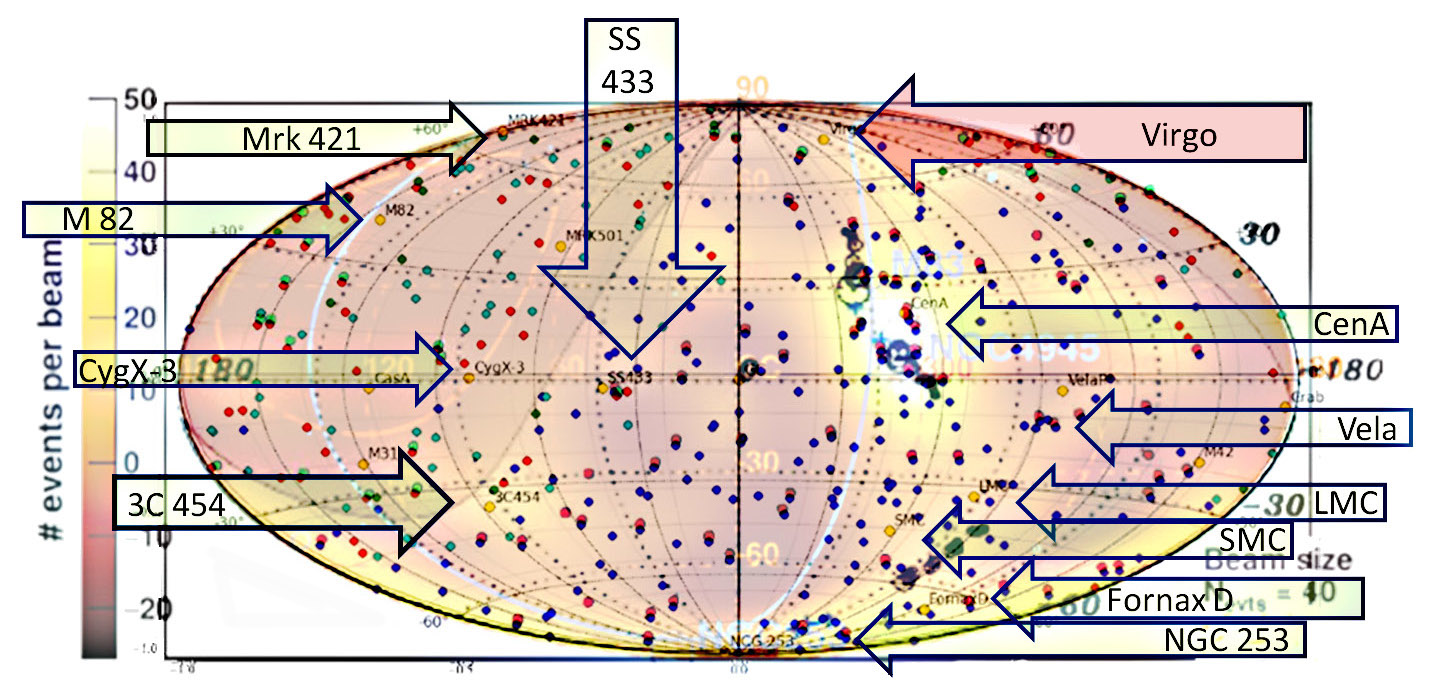}
    \caption{\label{Fargionfig8}}
  \end{subfigure}
  \hfill\mbox{}
  \caption{(a) This galactic UHECR map with in background the recent $39$ EeV anisotropy map, shows the earliest (2011) UHECR events and the 20 EeV smeared multiplet, up-down respect to the Cen~A position. We interpret these vertical spread of events as due to the planar galactic spiral magnetic fields, whose spin flip during the UHECR flight, leading to an up-down deflection   of the charged He-like UHECR \cite{fargion2011coherent}.\\
   (b) In present galactic coordinate sky map we recalled and tagged all the main probable sources (or the main missing ones, as Virgo  or eventual far AGN source, as 3C 454)\cite{Fargion2015Meaning}.
 A $39$ EeV anisotropy and a $20$ EeV multiplet map is in the background.
 These candidate sources  are not different from previous ones discussed since 2008-2015. In particular we like to note the very growing signals around the Vela nearby sky, the weak LMC-SMC  clustering multiplet  at twenty EeV, the enhanced NGC~253 and Fornax~D area of events at $40-60$ EeV.
 In addition to the 3C~454, also far Mrk~421 and possibly the far AGN PKS 0208-512 might also play a role as sources of UHECR clustering, if  their UHE ZeV neutrino courier, hitting the relic cosmic neutrino background with mass (possibly at $\simeq 1.6 eV$ as the recent candidate sterile one) are leading to nucleon and antinucleon secondaries observed as first UHECR cluster. }
\end{figure}

Now we consider the last individual  events over the last averaged anisotropy UHECR for highest $60$ EeV maps \cite{PAO(2018)}, (see Fig.\ref{Fargionfig8} small map in left corner). In this higher energy the UHECR clusterings are more narrow and clearer, as we should expect from the deflection law.
There are  growing evidences  that favour Cen~A AGN as a main source  while NGC~253 signal appears to be affirming itself also as a good source candidate  in Auger sky. In the TA north sky M82 it is a possible, although off axis source.
There are weaker clusterings in different areas.
There it is a most relevant one, below the Vela source, and a very peculiar clustering
toward the far AGN 3C~454, see Fig.\ref{Fargionfig8}.
The very recent Auger clustering at $60$ EeV \cite{PAO(2018)} is shown in an averaged color map, see in left corner of the Fig.~\ref{Fargionfig8}.
Here we show the most certain (NGC~253, Cen~A, M82) as well as the most probable galactic or extragalactic names and locations and galactic coordinates.

\begin{figure}[h]
  \mbox{}\hfill  
  \begin{subfigure}[t]{.48\linewidth}
    \centering\includegraphics[height=5cm, width=\linewidth]{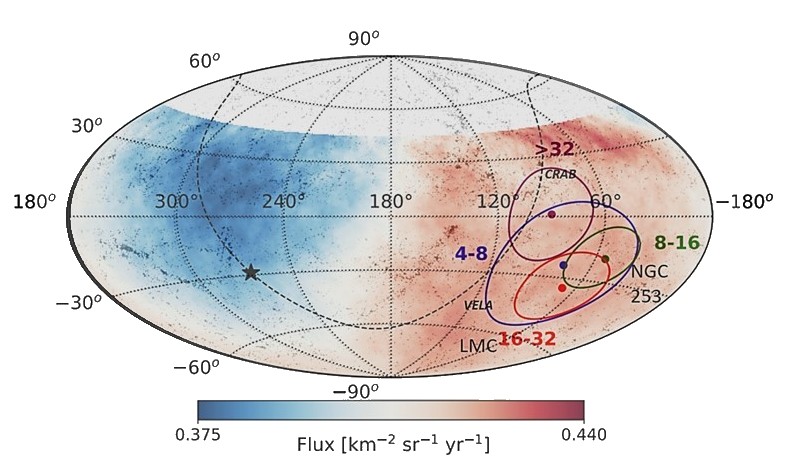}
    \caption{\label{dipole2c}}
  \end{subfigure}\hfill
  \begin{subfigure}[t]{.48\linewidth}
    \centering\includegraphics[height=5cm, width=\linewidth]{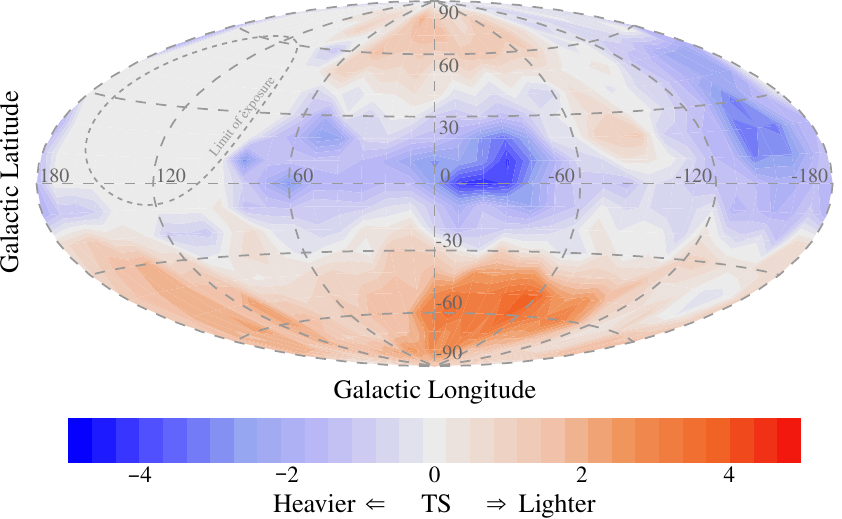}
    \caption{\label{Image31}}
  \end{subfigure}
  \hfill\mbox{}
  \caption{(a) The  UHECR dipole clustering directional rings at different energy ranges in equatorial coordinates. Few suggested galactic or nearby extragalactic source candidate are labeled.
  (b) The map in galactic coordinates, are derived for UHECR whose ⟨Xmax⟩ of the  event within 30 degrees for each bin to the rest of the sky.  The map colours are in Red for lightest (Blue for heaviest) UHECR mass anisotropy. Such signature suggest a He-like extragalactic component and a Ni-Co-Fe galactic one, possibly along nearest Crab-Vela galactic sources. These anisotropies are somehow complementary to the previous dipole ones.  \cite{Mayotte2022Aq}.}
\end{figure}

\section*{Summary}

UHECR Light or lightest nuclei are the main signature that guarantees two important results: Virgo UHECR cluster are obscured (as being  absent in AUGER and TA), the UHECR are in a very restricted local universe of just a few Mpc. This facilitated our identification of the UHECR possible locations tuned with the few narrow scale anisotropy. The twin hot spot in North and South sky can be associated to a few nearby AGN or starburst sources shown in previous figures and earlier references.
The 10 EeV dipole anisotropy cannot be  ruled by any far( hundred Mpc) extragalactic origin. Indeed if UHECR were made by proton reaching us from  such cosmic  volumes, their smeared tracks should lead to isotropy, while tens Mpc trace should point at least to, unobserved,  Virgo (or even a Great Attractor) cluster. We underline that the celebrated kinetic Doppler cosmic anisotropy is just at $ 2\cdot 10^{-3}$ versus the present huge  UHECR dipole, more than 20 time larger. No such high kinetic motion explanation is possible.   Neither any such mass anisotropy would be around  far (tens Mpc) universe. Indeed the Auger UHECR maps at few up to tens EeV have just shown the dipole in an very different direction of the sky. These volumes contains mainly nearest recent explosive galactic events as, Vela, Crab or near  LMC, NGC~253 or M31, (our nearest twin companion Andromeda galaxy); see.\cite{fargion2018JPSCP}. These sources  might be the main responsable ones for such dipole imprint.
Cen A, the brightest AGN in the hot spot, was confirmed in last  years as well as the M82 AGN,  NGC 253, and maybe, we claim here, our  M31: \cite{2015arXiv151208794F}. The nearest micro-quasars or early GRB jets \cite{Fargion(1999)} (as SS433 and  Vela and Crab gamma pulsar) may play a complementary  role, with LMC and SMC, at the ten EeV UHECR anisotropy. 
Few rare, non galactic clusterings, as the one around 3C~454, if confirmed, might be originated by the UHE ZeV neutrino scattering onto relic cosmic ones, with a light (even a sterile) eV size mass.  Time and growing data they will confirm or will wash away this exciting tuned astro-particle solution.
To disentangle such a possibility, one  might test of the UHECR composition study (by its $X_{max}$ or slant depth, see \cite{Gaisser1977Reliability} )  for each individual   arrival direction sources or map, \cite{Mayotte2022Aq}: while CenA, M82, NGC~253,  are here suggested to be lightest  nuclei events, the eventual ones, by ZeV neutrino scattering, versus Mrk421, 3C~454,  UHECR events, should  be a nucleon and anti-nucleon signals. 

\section{Conclusions}
In conclusion UHECR are very probably formed in AGN quasars and in micro-quasars (star forming regions, accretion disk and ejecta), powered by black Hole or neutron star binary tidal disruptions. These jets may arise in both AGN and in smaller GRB, micro quasars, and SGR, shining at different epochs and locations \cite{fargion1999nature}.\\
  The charged hadronic heavy nuclei UHECR from those jets are possibly bent, turned and even contained in spirals within their own source galaxy. However, their fragment secondaries at $40-60$ EeV, mostly Helium-like, are more directional and able to escape the AGN galaxy (Cen A, NGC 253, M82, ...),  reaching us with some memory of their origin, the hot spots.\\
  Protons from a hundred Mpcs and at few EeV can also escape galaxies. Their overlapping and bending should however lead to an isotropic sky.   
  The lightest UHECR nuclei at tens EeV  can fly only within a few Mpc, hiding the Virgo presence. 
  Lastly, the heaviest and most energetic Ni-Co-Fe UHECR are mainly bent, even contained as galactic traces, possibly by nearest sources as Crab, Vela or LMC.  Few unexplained UHECR correlation with far AGNs (Mrk, 3C 454, ...), if confirmed, may require a ZeV neutrino scattering on relic ones, overcoming any GZK cut-off.
  
   The few hardest and heavy UHECR, Ni-Co-Fe, can be partially originated in our own galaxy \cite{Mayotte2022Aq}, being even radiactive \cite{fargion2012apart}. The UHECR  anisotropy \cite{Mayotte2022Aq} and asymmetry \cite{plotko2022indication}, show their origin within a very local Universe \cite{fargion2018JPSCP}, in different times and places, as proved by several hints and signatures. More data will test, confirm or disprove the present reading keys.
\subsection*{Acknowledgement}
{\footnotesize The research by M.K. is financially supported by Southern Federal
University, 2020 Project VnGr/2020-03-IF.}

\FloatBarrier

\section*{References}
\bibliographystyle{iopart-num}
\bibliography{daf2022}

\end{document}